\documentstyle[12pt]{article}

\newcommand{\vs}[1]{\vspace{#1 mm}}

\newcommand{\beq}{\begin{equation}}
\newcommand{\eeq}{\end{equation}}
\newcommand{\beqa}{\begin{eqnarray}}
\newcommand{\eeqa}{\end{eqnarray}}

\newcommand{\eq}[1]{(\ref{#1})}

\newcommand{\eps}{\epsilon}

\newcommand{\ra}{\rightarrow}
\newcommand{\del}{\partial}
\newcommand{\NP}[1]{Nucl.~ Phys.~ {\bf #1}}
\newcommand{\PL}[1]{Phys. ~Lett.~ {\bf #1}}

\newcommand{\PTP}[1]{Prog.~ Theor.~ Phys.~ {\bf #1}}

\newcommand{\MPL}[1]{Mod.~ Phys.~ Lett.~ {\bf #1}}

\begin{document}
\topmargin 0pt
\oddsidemargin 5mm
\begin{titlepage}
\setcounter{page}{0}
\begin{flushright}
  NBI-HE-95-34 \\
  October, 1995
\end{flushright}
\vs{13}
\begin{center}
{\large  OPE formulae for deformed super-Virasoro algebras}
\vs{20}

Haru-Tada Sato
\footnote{Fellow of Danish Research Academy\\
\mbox{}\hspace{0.4cm} E-mail address : sato@nbivax.nbi.dk}

\vs{5}

{\em The Niels Bohr Institute, University of Copenhagen\\
Blegdamsvej 17, DK-2100 Copenhagen, Denmark}
%
%

\end{center}
\vs{10}

\begin{abstract}
We show the OPE formulae for three types of deformed super-Virasoro
algebras: Chaichian-Presnajder's deformation, Belov-Chaltikhian's one
and its modified version. Fundamental (anti-)commutation relations
toward a ghost realization of deformed super-Virasoro algebra are also
discussed.
\end{abstract}

\vspace{1cm}

\end{titlepage}
\newpage
%
%
\section{Introduction}
\indent

 In this paper, we analyze the deformed super-Virasoro superalgebras, based
on the Sugawara construction, proposed in recent years \cite{CP},\cite{BC}.
It might be useful for generalizing conformal field and string theories to
investigate the deformed super-Virasoro algebras, which are different from
$W$-infinity algebras in the sense of including an independent parameter,
namely, a deformation parameter. For example, in conformal field theories,
exactly speaking in the thermodynamic limit of the XYZ Heisenberg chain model
\cite{XYZ}, there exists a deformed Virasoro algebra \cite{LP},\cite{DVA1},
\cite{DVA2} which satisfies quadratic relations related to the
Zamolodchikov-Faddeev algebra for deformed chiral vertex operators. The
algebra of \cite{CP} corresponds to a particular case of this case \cite{LP}.
As for deformation of string theories, it is interesting to analyze the
deformed Virasoro algebra on a speculative ground that the Virasoro
constraints would lead to deformation of the BRS operator, conformal
anomalies, and thus to the critical dimensions depending on a deformation
parameter $q$. The $q$ might be related to something like an energy scale
where the spacetime dimension changes.

 There are two types of the deformed Virasoro algebras which are made by the
Sugawara construction, namely in terms of the bosonic and fermionic Heisenberg
oscillators \cite{CP}. Each possesses different structure constants but
reduces to the same Virasoro algebra in $q\ra1$ up to the difference of
central charges. We refer to the bosonic realization as $q$-${\rm Vir}^B$ and
to the fermionic one as $q$-${\rm Vir}^F$ for convenience (this naming follows
from \cite{BC}). Further, Chaichian and Presnajder introduced a deformed
spin-3/2 current and proposed a deformed supersymmetric Virasoro algebra
which is constituted of $q$-${\rm Vir}^F$, $q$-${\rm Vir}^B$ and the deformed
superconformal current. However much inconvenience will be raised by the
difference of structure constants between bosonic and fermionic algebras in:
anomaly cancellation, superfield formulation, deformation of the ghost
Virasoro and BRS algebras.

Belov and Chaltikhian improved the unpleasant feature of the above deformed
super-Virasoro algebra and proposed a supersymmetric $q$-${\rm Vir}^B$ algebra.
They modified the basis of the fermionic Heisenberg algebra in order that the
fermionic $q$-Virasoro generators satisfy the same commutation relation as
$q$-${\rm Vir}^B$. This achievement means that the fermionic and bosonic
sectors can be organized into unique commutation relation or operator product
expansion (OPE), which is convenient to construct BRS operator and to discuss
other algebraic structure.  We should furthermore find ghost realization of
the deformed Virasoro algebra in order to make a progress in deformed string
theories. To find a complete supersymmetric deformed Virasoro algebra, it is
thus useful to list up OPE formulae for several types of supersymmetric
deformed Virasoro algebra.

In sect. 2, we review Chaichian-Presnajder's superalgebra adding
brief comments on the Ramond sector. Then, we provide the OPE formulae for
this superalgebra. In sect. 3, we are concerned with Belov-Chaltikhian's
superalgebra in which they have not shown any (anti-)commutators other than one
of (anti-)commutation relations. We first complete all (anti-)commutators
including central extensions for this superalgebra and then exhibit the OPE
formulae for them. Simplification of this superalgebra is also presented.
In sect. 4, we intend to discuss the deformed Virasoro algebras in terms of
$bc$ system. New type of deformed Virasoro algebra, a realization of
$q$-${\rm Vir}^F$ and basic set of algebras toward a $q$-Virasoro superalgebra
are shown. Conclusion is in sect. 5.

\setcounter{equation}{0}
\section{Chaichian-Presnajder's superalgebra}
\indent

Let us first explain some notations which make expressions simple. We
introduce the following $q$-brackets for an arbitrary rational number $x$:
\beq
[x]={q^x-q^{-x} \over q-q^{-1}}\,\quad\quad
[x]_{\pm}=(q^x \pm q^{-x})/2,
\eeq
where $q$ is not a root of unity to avoid $[x]=0$ for any $x$.
The $q$-difference of $f(z)$ at $z=a$ is
\beq
\del^q f(a) = \del_z^qf(z)\vert_{z=a}={f(zq) - f(zq^{-1}) \over
z(q-q^{-1})}\vert_{z=a}
\eeq
and note that the difference $\del^q f(za)\not=\del^q_z f(za)$.
It is also convenient to use
\beq (z-w)^2_q = (z-qw)(z-q^{-1}w). \eeq
Finally, we often use the abbreviation for two letters $k$ and $l$ for
simplicity
\beq Q=q^{k/2} \hskip 20pt \mbox{and} \hskip 20pt P=q^{l/2}. \eeq

Chaichian and Presnajder proposed the following deformation of the $N=1$
super-Virasoro algebra  \cite{CP}:
\beq \hskip -10pt
[F(n,k),F(m,l)]=\sum_{\eps=\pm 1} {[(n\eps l-mk)/2][k+\eps l]
                \over [k][\eps l]} F(n+m,k+\eps l)
                + C^F(n\vert k,l)\delta_{n+m,0}               \label{FF}
\eeq
\beq
[B(n,k),B(m,l)]={1\over2}\sum_{\eps,\eta=\pm 1}
                [{n(\eta l+1)-m(\eps\eta k+1)\over2}] B(n+m,\eps k+l+\eta)
                + C^B(n\vert k,l)\delta_{n+m,0}                 \label{BB}
\eeq
\beq \hskip -125pt
[F(n,k),G(r,l)]={1\over[k](q-q^{-1})}\sum_{\eps=\pm 1}
                \eps q^{(nl-\eps kr)/2} G(n+r,l+\eps k)        \label{FG}
\eeq
\beq \hskip -75pt
[B(n,k),G(r,l)]={-1\over2(q-q^{-1})}\sum_{\eps,\eta=\pm 1} \eta
         q^{{r(\eps k+\eta)-n(1+\eta)\over2}} G(n+r,\eps k+l+\eta) \label{BG}
\eeq
\[  \hskip -130pt
\{G(r,k),G(s,l)\}=  C^G(r\vert l,k)\delta_{r+s,0} + 2q^{(rl+sk)/2}B(r+s,k-l)\]
\beq           +\sum_{\eps=\pm 1}\eps
              q^{r(\eps-l)-s(k+\eps)\over2}[k-l+\eps]F(r+s,k-l+\eps),\label{GG}
\eeq
where
\beq \hskip -5pt
C^F(n\vert k,l)={1\over2[k][l]}\sum_{m=1}^n
                [({n+1\over2}-m)k][({n+1\over2}-m)l]\eeq
\beq
C^B(n\vert k,l)={1\over2}\sum_{m=1}^n
                [({n\over2}-m)k]_+[({n\over2}-m)l]_+[m][n-m]\eeq
\beq \hskip -1pt
C^G(r\vert l,k)=\sum_{m=1}^{r+{1\over2}}q^{(m-{r+1\over2})(l-k)}[r+{1\over2}-m]
=C^G(-r\vert k,l).\eeq
This set of commutators reduces to the usual $N=1$ super-Virasoro algebra in
$q\ra1$. Eq.\eq{FF} is called $q$-${\rm Vir}^F$ and \eq{BB} is
$q$-${\rm Vir}^B$ in the reference \cite{BC} for the reason that they are
realized by the following Sugawara constructions
\beq
F(n,k)={1\over2[k]}\sum_{r\in{Z+1/2}} [({n\over2}-r)k]:b_r b_{n-r}:
\eeq
\beq \hskip -5pt
B(n,k)={1\over2}\sum_{m\in Z} [({n\over2}-m)k]_+ :a_m a_{n-m}:
\eeq
\beq \hskip -30pt
G(r,k)=\sum_{m\in Z} q^{k(m-r/2)}:a_m b_{r-m}:
\eeq
with the deformed boson $a_n$ and the Neveu-Schwarz fermion $b_r$ defined by
\beq
 [a_n,a_m]=[n]\delta_{n+m,0}, \hskip 40pt \{b_r,b_s\}=\delta_{r+s,0}.
\eeq
Note that $F$ and $B$ are symmetric under $k\ra-k$ but $G$ is not.
The realizations of $q$-${\rm Vir}^F$ and $q$-${\rm Vir}^B$ in terms of
differential operators are given in \cite{saito} and \cite{KS}. On the other
hand for the supercurrent $G$, it has not been given.

Before going over the OPE formulae for the above algebra, let us notice on the
Ramond sector. For the Ramond fermion, we have only to change the
half-integer index $r$ stemming from the fermion into integer one. Using
\beq
[F(n,k),b_m]=-{1\over[k]}[({n\over2}+m)k]b_{n+m},\quad
[B(n,k),a_m]=-[m][({n\over2}+m)k]_+ a_{n+m} \eeq
\beq \hskip -60pt
\{G(n,k),b_m\}=q^{({n\over2}+m)k} a_{n+m},\quad\hskip 43pt
[G(n,k),a_m]=-[m]q^{-({n\over2}+m)k} b_{n+m},\eeq
we can check that the same commutation relations as \eq{FF}-\eq{GG} are
satisfied with the central extensions $C^F$ and $C^G$ replaced with
\beq \hskip -88pt
C^F_R(n\vert k,l)={1\over2[k][l]}\sum_{m=1}^n
                  [({n\over2}-m)k][({n\over2}-m)l]\eeq
\beq \hskip -60pt
C^G_R(n\vert l,k)={1\over2}q^{-n(l-k)/2}[n]+\sum_{m=1}^{n}
             q^{(m-{n\over2})(l-k)}[n-m] = C^G_R(-n\vert k,l).\eeq
The difference between two sectors is only constant shift between the zero
mode of $F$ generators
\beq
F(n,k)_{NS} = F(n,k)_R + {[k/4]^2\over2[k][k/2]}\delta_{n,0}.
\eeq
We accordingly concentrate on the NS sector in the following argument.

Now we incorporate the above generators into the
following composite fields as the Fourier mode operators:
\beq
F(z;k)=\sum_n F(n,k)z^{-n-2}={-1\over Q+Q^{-1}}:\psi(zQ^{-1})\del^Q \psi(z):
\eeq
\beq \hskip -35pt
B(z;k)=\sum_n B(n,k)z^{-n-2}={1\over2}:\Phi(zQ)\Phi(zQ^{-1}):
\eeq
\beq \hskip -10pt
G(z;k)=\sum_r G(r,k)z^{-r-3/2}=Q^{-1/2}:\psi(zQ)\Phi(zQ^{-1}):, \label{gfield}
\eeq
where
\beq \Phi(z)=\sum_n a_n z^{-n-1},\hskip 40pt
      \psi(z)=\sum_r b_r z^{-r-1/2}. \eeq
Singular parts of these bosonic and fermionic fields are expressed as
\beq
\psi(z)\psi(w)={1\over z-w}, \hskip 40pt \Phi(z)\Phi(w)={1\over(z-w)^2_q},
\eeq
and then
\beq
G(z;k)\psi(w)={Q^{-1/2}\over zQ-w}\Phi(wQ^{-2})
\eeq
\beq
G(z;k)\Phi(w)=Q^{-1/2}\del^q_w\left( {\psi(wQ^2)\over zQ^{-1}-w} \right)
\eeq
\beq \hskip -38pt
F(z;k)\psi(w)={1\over Q+Q^{-1}}\left\{ {\psi(w)\over (z-w)^2_Q}
+{\del^Q\psi(wQ)\over zQ^{-1}-w}+{\del^Q\psi(wQ^{-1})\over zQ-w} \right\}
\eeq
\beq
B(z;k)\Phi(w)={1\over2}\left\{ {\Phi(wQ^{-2}q^{-1})\over(zQ-w)^2_q}
+{Q^{-2}\del^q\Phi(wQ^{-2})\over zQ-wq}+\quad (\,k \ra -k\,)\right\}.
\eeq
With these OPE formulae, we obtain the OPEs corresponding to \eq{FF}-\eq{GG}
up to regular terms
\[ \hskip -119pt
F(z;k)F(w;l)={1\over(Q+Q^{-1})(P+P^{-1})(z-wP)^2_Q(z-wP^{-1})^2_Q} \]
\beq  \hskip 20pt  + {1\over w(q-q^{-1})}\sum_{\eps,\eta=\pm1}
             {[\eps k+\eta l]\over[\eps k][\eta l]}
          {P^{-\eta}\over zQ^{-\eps}-wP^{\eta}}F(wQ^\eps;\eps k+\eta l)
\label{opeFF}
\eeq
\[ \hskip -169pt
B(z;k)B(w;l)=\sum_{\nu=\pm1}{1/4\over(zQ-wP^\nu)^2_q(zQ^{-1}-wP^{-\nu})^2_q}\]
\beq  \hskip 60pt   +{1\over w(q-q^{-1})}\sum_{\eps,\eta,\nu=\pm1}
                    {\eta P^{-\nu}/2\over zQ^{-\eps}-wP^{\nu}q^\eta}
                    B(wQ^\eps q^{\eta/2};\eps k+\nu l+\eta)       \label{opeBB}
\eeq
\beq \hskip -124pt
F(z;k)G(w;l)={P^{-1}\over w[k](q-q^{-1})}\sum_{\eps=\pm1}
          {\eps Q^{\eps/2}\over z-wPQ^\eps}G(wQ^\eps;\eps k+l)   \label{opeFG}
\eeq
\beq \hskip -26pt
B(z;k)G(w;l)={P\over 2w(q-q^{-1})}\sum_{\eps,\eta=\pm1}
         {\eta (Q^\eps q^{\eta/2})^{-1/2} \over zQ^{-\eps}-wP^{-1}q^\eta}
         G(wQ^\eps q^{\eta/2};l-\eps k-\eta)                     \label{opeBG}
\eeq
\beq \hskip -23pt
G(z;k)G(w;l)={(PQ)^{-1/2}\over (zQ-wP)(zQ^{-1}-wP^{-1})^2_q}+
{2(PQ)^{-1/2}\over zQ-wP} B(wQ^{-1};l-k)                        \label{opeGG}
\eeq\[ +\sum_\eps{ \eps(PQq^\eps)^{1/2}\over zQ^{-1}-wP^{-1}q^\eps}[k-l+\eps]
F(wQq^{\eps/2};k-l+\eps).
\]
\eq{opeFF} and \eq{opeBB} coincide with those in refs.\cite{hsato} and
\cite{OS}. \eq{opeFG}-\eq{opeGG} are checked by evaluating
\beq
[F(n,k),G(r,l)]={1\over(2\pi i)^2}
               \oint_0dw\oint_Pdzz^{n+1}w^{r+1/2}F(z,k)G(w,l)
\eeq
\beq
[B(n,k),G(r,l)]={1\over(2\pi i)^2}
               \oint_0dw\oint_Pdzz^{n+1}w^{r+1/2}B(z,k)G(w,l)
\eeq
\beq
[G(r,k),G(s,l)]={1\over(2\pi i)^2}
               \oint_0dw\oint_Pdzz^{r+1/2}w^{r+1/2}G(z,k)G(w,l),
\eeq
where $\oint_P$ counts contribution from all poles in $z$ plane.
In appendix, we show an alternative representation of the OPEs written
in (anti-)commutator forms.

\setcounter{equation}{0}
\section{Belov-Chaltikhian's superalgebra}
\indent

In this section, we choose different normalization for the fermionic Heisenberg
algebra
\beq \{d_r,d_s\}=[r]_+\delta_{r+s,0}. \eeq
Belov and Chaltikhian showed that
\beq H(n,k)={1\over2}\sum_r [k({n\over2}-r)]:d_rd_{n-r}: \eeq
satisfies the $q$-${\rm Vir}^B$ algebra up to a central term for the NS fermion
and discussed a supersymmetric extension \cite{BC}. However they have shown
neither central extensions nor full set of supercurrent commutators.
In the following, we complete all (anti-)commutation relations with central
extensions including the Ramond fermion case. After that, we show the OPE
formulae for their superalgebra.

According to ref.\cite{BC}, we consider the following deformation of a
supercurrent
\beq G^a(r,k)=\sum_m[({r\over2}-m)k]_a :a_md_{r-m}:,\quad\quad(a=\pm)
              \label{charge}\eeq
and the sum of bosonic and fermionic realizations of the $q$-${\rm Vir}^B$
generators
\beq L(n,k)= H(n,k)+B(n,k). \eeq
In contrast with the Chaichian-Presnajder deformation,
$G^+$ is symmetric and $G^-$ and $H$ are antisymmetric. $L$ is
no longer symmetric in $k$. Commutation relations between the deformed
super-Virasoro generators and the Heisenberg oscillators are
\beq
[L(n,k),d_m]=-[m]_+[({n\over2}+m)k]d_{n+m},\quad
[L(n,k),a_m]=-[m][({n\over2}+m)k]_+ a_{n+m} \eeq
\beq \hskip -5pt
\{G^a(r,k),d_s\}=a[s]_+[({r\over2}+s)k]_a a_{r+s},\quad
\hskip 5pt [G^a(r,k),a_m]=-[m][({r\over2}+m)k]_a d_{r+m}\label{gd}.
\eeq
With use of these formulae, we obtain the following superalgebra including
central extensions
\[ \hskip -77pt
[L(n,k),L(m,l)]=\sum_{\eps,\eta=\pm1}{1\over2}[{n(\eps l+1)-m(\eta k+1)\over2}]
L(n+m,\eps k+\eta l+\eps\eta)\]\beq \hskip -60pt
+(C^H(n\vert k,l)+C^B(n\vert k,l))\delta_{n+m,0}                \label{belLL}
\eeq
\beq \hskip -11pt
[L(n,k),G^a(r,l)]={1\over q-q^{-1}}\sum_{\eps,\eta=\pm1}\eta^{1-a\over2}
                  [{n(\eta l+1)-r(\eps k+1)\over2}]_{-\eps}
                  G^{a\eps}(n+r,\eps k+\eta l+1)               \label{belLG}
\eeq
\[ \hskip -15pt
\{G^a(r,k),G^b(s,l)\}={1\over2}\sum_{\eps,\eta}(-\eps)^{1-a\over2}
(-\eta)^{1-b\over2}[{r(\eta l+ab)-s(\eps k+ab)\over2}]_{ab}
L(r+s,\eps k+\eta l+ab)\]\beq\hskip -20pt
+{1\over4}\sum_{\eps,\eta}(-\eps)^{1-a\over2}(-\eta)^{1-b\over2}
            {\tilde C}^G(s\vert \eps k,\eta l)\delta_{r+s,0},   \label{belGG}
\eeq
where $C^H(n\vert k,l)$ is
\beq
      {1\over2}\sum_{m=1}^n[({n+1\over2}-m)k][({n+1\over2}-m)l]
                [n-m+{1\over2}]_+[m-{1\over2}]_+, \quad \mbox{for NS}
\eeq
\beq
{1\over2}\sum_{m=1}^n[({n\over2}-m)k][({n\over2}-m)l] [n-m]_+[m]_+
\quad\quad \hskip 75pt\mbox{for R}
\eeq
and
\beq {\tilde C}^G(r\vert l,k)=\sum_{r\leq m<r+s}
             q^{({s\over2}+r-m)l+({r\over2}-m)k}[m-r][m]_+,
\phantom{OP} \left\{
\begin{array}{ll}
             m\in {\bf Z}+{1\over2}, &\quad \mbox{for NS} \\
             m\in {\bf Z},           &\quad \mbox{for R}
\end{array}\right..
\eeq
We should keep in mind that the number of (anti-)commutation relations is
larger than that of usual case, since we have a redundant degree of freedom
expressed by the superscript $a$.

OPEs can be calculated through the following field operators
\beq  {\tilde\psi}(z)=\sum_r d_r z^{-r-1/2}   \eeq
\beq   T(z;k)=B(z;k)+H(z;k)                   \eeq
\beq   G^a(z;k)={1\over2}\sum_\eps(-\eps)^{1-a\over2}
                {\tilde G}(z;\eps k),           \eeq
where
\beq
H(z;k)={1\over z(q-q^{-1})}{\tilde\psi}(zQ){\tilde\psi}(zQ^{-1})
\eeq
and ${\tilde G}(z;k)$ is defined by the replacement $\psi\ra{\tilde\psi}$
in eq.\eq{gfield}. Singular parts of a product of ${\tilde\psi}$ is
\beq
{\tilde\psi}(z){\tilde\psi}(w)={1\over2}({q^{1/2}\over z-wq}
                                        +{q^{-1/2}\over z-wq^{-1}})
\eeq
and then
\beq {\tilde G}(z;k)\Phi(w)=Q^{-1/2}\partial_w^q
\left({ {\tilde\psi}(wQ^2)\over zQ^{-1}-w}\right), \hskip 15pt
{\tilde G}(z;k){\tilde\psi}(w)={1\over2}\sum_\eps{Q^{-1/2}q^{\eps/2}\over
zQ-wq^\eps}\Phi(wq^\eps Q^{-2}),
\eeq
\beq
T(z;k){\tilde\psi}(w)=H(z;k){\tilde\psi}(w)={1\over2w(q-q^{-1})}\sum_{\eps,\nu}
{-\eps q^{\nu/2}Q^\eps\over
zQ^\eps-wq^{-\nu}}{\tilde\psi}(wq^{-\nu}Q^{-2\eps}).
\eeq
We finally obtain the following OPE formulae (NS sector)
\[ \hskip -140pt
T(z;k)T(w;l)={1\over2w(q-q^{-1})}\sum_{\eps,\nu,\eta}\eta
{T(wQ^{\eps}q^{\eta/2};\eps\eta l+\nu\eta k+\eps\nu)\over
(zQ^{-\eps}-wP^\nu q^\eta)P^\nu}    \]
\beq \hskip 30pt
+{[l/2]/4\over q-q^{-1}}\sum_{\eps,\nu}{q^\eps Q-q^\nu Q^{-1}\over
(zQ-wq^\nu)^2_P(zQ^{-1}-wq^\eps)^2_P}+
\sum_{\nu=\pm1}{1/4\over(zQ-wP^\nu)^2_q(zQ^{-1}-wP^{-\nu})^2_q}
\eeq
\beq \hskip -10pt
T(z;k)G^a(w;l)={1\over2w(q-q^{-1})}\sum_{\eps,\eta,\nu}
{\eps\eta\nu^{1-a\over2}(Q^\eps q^{\eps\eta/2})^{-1/2}\over
(zQ^{-\eps}-wP^\nu q^{\eta\eps})P^\nu}G^{\eta a}
(wQ^\eps q^{\eps\eta/2};\eps k+\nu l+\eps\eta)    \label{xx}
\eeq
\[ \hskip -88pt
G^a(z;k)G^b(w;l)={1\over4}\sum_{\eps,\eta,\nu}(-\eps)^{1-a\over2}
                     (-\nu)^{1-b\over2}
       {(P^\nu Q^\eps q^{-\eta})^{-1/2}\over zQ^\eps-wP^\nu q^\eta}
        \Bigl({1/2\over(zQ^{-\eps}-wP^{-\nu})^2_q}\]\beq \hskip -60pt
      +T(wQ^{-\eps}q^{\eta/2};ab\eta\nu l-ab\eta\eps k+ab) \Bigr). \label{yy}
\eeq
These results are checked in the similar way as previous section.

Finally, we make a comment on this superalgebra. Although we have
introduced two kind of deformed supercurrents \eq{charge} following to
ref.\cite{BC}, we can simplify the algebra and reduce the number of
anti-commutation relations among $G^{\pm}$. Decomposing $G^{\pm}$ into two
pieces as
\beq
G^\pm(r,k)={1\over2}({\tilde G}(r,-k)\pm {\tilde G}(r,k)), \label{simpleG}
\eeq
we change eq.\eq{gd} to
\beq \{{\tilde G}(r,k),d_s\}=q^{({r\over2}+s)k}[s]_+ a_{r+s},\quad
[{\tilde G}(r,k),a_m] = -[m]_+ q^{-({r\over2}+s)k} d_{r+m}.
 \eeq
Commutation relations among $B$, $H$ and ${\tilde G}$ are in turn
\beq  [\,B(n,k),{\tilde G}(r,l)\,]=\quad G \ra {\tilde G} \quad {\rm in}
\quad\eq{opeBG} \eeq
\beq
[\,H(n,k),{\tilde G}(r,l)\,]={1\over2(q-q^{-1})}\sum_{\eps,\eta}\eps
q^{n(l+\eta)-r(\eps k+\eta)\over2}{\tilde G}(n+r,l+\eps k+\eta)
\eeq
\[ \hskip -90pt
\{{\tilde G}(r,k),{\tilde G}(s,l)\}=
\sum_\eps\Bigl(q^{sk+rl+(s-r)\eps}B(r+s,k-l+\eps)\]
\beq   +\eps q^{-sk-rl-(s-r)\eps}H(r+s,k-l+\eps)\Bigr)
       +{\tilde C}^G(r\vert l,k)\delta_{r+s,0}.
\eeq
Taking account of the properties
\beq    B(n,k)=B(n,-k), \quad\quad  H(n,k)=-H(n,-k), \eeq
we can cast them into the following simple superalgebra
\beq
[L(n,k),{\tilde G}(r,l)]={1\over(q-q^{-1})}\sum_{\eps,\eta}\eps
[{n(l+\eta)-r(\eps k+\eta)\over2}]_{-\eps\eta}{\tilde G}(n+r,l+\eps k+\eta)
\label{LGtil}
\eeq
\beq
\{{\tilde G}(r,k),{\tilde G}(s,l)\}=
            \sum_{\eps,\eta}[{s(k+\eps)+r(l-\eps)\over2}]_{\eps\eta}
        L(r+s,\eta k-\eta l+\eps\eta)
        +{\tilde C}^G(r\vert l,k)\delta_{r+s,0}.         \label{GtilGtil}
\eeq
Here the number of independent (anti-)commutator brackets in
\eq{belLL}-\eq{belGG} reduces to 3 from 6. $L$ and ${\tilde G}$ are now of
course not symmetric under $k\ra -k$. OPEs for \eq{LGtil} and
\eq{GtilGtil} are given by
\beq
T(z;k){\tilde G}(w;l)={1\over2w(q-q^{-1})}\sum_{\eps,\nu,\eta}
{\eps^{1+\nu\over2}\eta^{1-\nu\over2}(Q^\eps q^{\eta/2})^{-1/2}\over
(zQ^{-\eps}-wP^\nu q^\eta)P^\nu}
     {\tilde G}(wQ^\eps q^{\eta/2};l+\eps\nu k+\nu\eta)
\eeq
\[ \hskip -50pt
{\tilde G}(z;k){\tilde G}(w;l)={1\over2}\sum_{\eps,\nu,\eta}
\Bigl((-\eps\eta)^{1-\nu\over2}{(P^\nu Q^\nu q^{-\eps})^{-1/2}\over
zQ^\nu-wP^\nu q^\eps}T(wQ^{-\nu}q^{\eps/2};\eta l-\eta k+\eps\nu\eta) \]\beq
+ {{1\over4}(PQq^{-\eps})^{-1/2}\over
(zQ-wPq^\eps)(zQ^{-1}-wP^{-1})^2_q}\Bigr).
\eeq
These are consistent with \eq{xx} and \eq{yy} under the relation \eq{simpleG}.

\setcounter{equation}{0}
\section{Deformation in $bc$ system}
\subsection{deformed algebra $q$-${\rm Vir}^{bc}$}
\indent

Let us consider the following deformed ghost Virasoro operators
\beq
L^{bc}(n,k)={1\over[k]}\sum_m[k(n\Delta-n+m)]:b_{n-m}c_m: \label{bc}
\eeq
with the commutation relation \cite{FMS}
\beq
 c_n b_m + \eps b_m c_n = \delta_{n+m,0}
\eeq
where $\Delta$ is the conformal weight for $b$ and $\eps$ means the statistics
($\Delta=2$, $\eps=1$ for ghosts and $\Delta=3/2$, $\eps=-1$ for superghosts)
and
\beq
b(z)=\sum_{n\in Z-\Delta}b_n z^{-n-\Delta}, \quad\quad
c(z)=\sum_{n\in Z+\Delta}c_n z^{-n-(1-\Delta)}.
\eeq
Eq.\eq{bc} satisfies
\beq
[L^{bc}(n,k),b_m]={1\over[k]}[k(n\Delta-n-m)]b_{n+m},\quad
[L^{bc}(n,k),c_m]={-1\over[k]}[k(n\Delta+m)]c_{n+m}.
\eeq
If we put $\Delta=1/2$, the generators \eq{bc} are similar to those for
$q$-${\rm Vir}^F$ and actually satisfy the same commutation relation as
\eq{FF} with the central term $2C^F$. While for any other values of $\Delta$,
we have to introduce an additional set of operators which vanishes in the
limit $q\ra1$ in order that its algebra may be closed. For example, we choose
it
\beq
O^{bc}(n,k)={[1]_-\over[k]}\sum_m[k(n\Delta-n+m)]_+:b_{n-m}c_m:,\label{Obc}
\eeq
which satisfies
\beq
[O^{bc}(n,k),b_m]={[1]_-\over[k]}[k(n\Delta-n-m)]_+b_{n+m},\quad
[O^{bc}(n,k),c_m]={-[1]_-\over[k]}[k(n\Delta+m)]_+c_{n+m}.
\eeq
We thereby obtain the following closed algebra as a deformed Virasoro algebra
(for later convenience, we call it $q$-${\rm Vir}^{bc}$)
\[
[L^{bc}(n,k),L^{bc}(m,l)]=\sum_{\eps=\pm 1} {[(n\eps l-mk)/2][k+\eps l]
                \over [k][\eps l]}
\Bigl( [({1\over2}-\Delta)(\eps nl+mk)]_+L^{bc}(n+m,k+\eps l)\]\beq
+[({1\over2}-\Delta)(\eps nl+mk)] O^{bc}(n+m,k+\eps l)/[1]_-\Bigr)
+C^{LL}(n\vert k,l)\delta_{n+m,0},\label{bcbc}
\eeq
\[
[L^{bc}(n,k),O^{bc}(m,l)]=\sum_{\eps=\pm 1} {[(n\eps l-mk)/2][k+\eps l]
                \over [k][l]}
\Bigl( [({1\over2}-\Delta)(\eps nl+mk)]_+O^{bc}(n+m,k+\eps l)\]\beq
 +[1]^2_-[({1\over2}-\Delta)(\eps nl+mk)]_- L^{bc}(n+m,k+\eps l)\Bigr)
+C^{LO}(n\vert k,l)\delta_{n+m,0},
\eeq
\[
[O^{bc}(n,k),O^{bc}(m,l)]=\sum_{\eps=\pm 1} {[(n\eps l-mk)/2][k+\eps l]
                \over [k][l]}
\Bigl([1]^4_- [({1\over2}-\Delta)(\eps nl+mk)]_+L^{bc}(n+m,k+\eps l)\]\beq
+[1]^2_-[({1\over2}-\Delta)(\eps nl+mk)]_- O^{bc}(n+m,k+\eps l)\Bigr)
+C^{OO}(n\vert k,l)\delta_{n+m,0}
\eeq
where
\beq
C^{LL}(n\vert k,l)={\eps\over2[k][l]}(\sum_{0<m\leq n}
+\sum_{0\leq m<n})_{m\in Z-\Delta} [(n\Delta-m)k][(n-m-n\Delta)l],\eeq
\beq
C^{LO}(n\vert k,l)={\eps[1]_-\over2[k][l]}(\sum_{0<m\leq n}
+\sum_{0\leq m<n})_{m\in Z-\Delta} [(n\Delta-m)k][(n-m-n\Delta)l]_+,\eeq
\beq
C^{OO}(n\vert k,l)={\eps[1]^2_-\over2[k][l]}(\sum_{0<m\leq n}
+\sum_{0\leq m<n})_{m\in Z-\Delta} [(n\Delta-m)k]_+[(n-m-n\Delta)l]_+.\eeq
Only \eq{bcbc} survives in $q\ra1$ reducing to the usual ghost Virasoro
algebra. Note that the center $C^{LL}$ coincides with $2C^F_R$ for sum over
integers on $m$ as well as with $2C^F$ for half integers after putting
$\Delta=1/2$ and $\eps=1$. This is very similar to the $q=1$ case.

\subsection{realization of $q$-${\rm Vir}^F$ }
\indent

The set of eqs.\eq{bc} and \eq{Obc} is composed of a linear combination of
two independent (positive/negative) powers of $q$, namely of
$\sum q^mb_{n-m}c_m$ and $\sum q^{-m}b_{n-m}c_m$. Recombining these operators
we can eliminate the extra operator $O^{bc}$ from \eq{bcbc}, and we thus
obtain a closed algebra for single generator. To see this, let us consider
the following operators as a recombination:
\beq
F^{bc}(n,k)={-1\over[k]}\sum_m[k({n\over2}-m)]:b_{n-m}c_m: \label{Fbc}
\eeq
\beq
R^{bc}(n,k)={1\over[k]}\sum_m[k({n\over2}-m)]_+:b_{n-m}c_m:, \label{Rbc}
\eeq
which satisfies
\[
[F^{bc}(n,k),b_m]={-1\over[k]}[k({n\over2}+m)]b_{n+m},\quad
[R^{bc}(n,k),b_m]={1\over[k]}[k({n\over2}+m)]_+b_{n+m}
\]
\beq
[F^{bc}(n,k),c_m]={1\over[k]}[k({n\over2}+m)]c_{n+m},\quad
[R^{bc}(n,k),c_m]={-1\over[k]}[k({n\over2}+m)]_+c_{n+m}.
\eeq
With these formulae, we obtain the following results
\beq
[F^{bc}(n,k),F^{bc}(m,l)]=\sum_{\eps=\pm 1} {[(n\eps l-mk)/2][k+\eps l]
                \over [k][\eps l]}F^{bc}(n+m,k+\eps l)
                +C^{FF}(n\vert k,l)\delta_{n+m,0}  \label{FFbc}
\eeq
\beq
[F^{bc}(n,k),R^{bc}(m,l)]=\sum_{\eps=\pm 1} {[(n\eps l-mk)/2][k+\eps l]
                \over [k][l]}R^{bc}(n+m,k+\eps l)
                +C^{FR}(n\vert k,l)\delta_{n+m,0}
\eeq
\beq
[R^{bc}(n,k),R^{bc}(m,l)]=[1]_-^2\sum_{\eps=\pm 1}
{[(n\eps l-mk)/2][k+\eps l] \over [k][l]} F^{bc}(n+m,k+\eps l)
+C^{RR}(n\vert k,l)\delta_{n+m,0},
\eeq
where
\beq
C^{FF}(n\vert k,l)={\eps\over[k][l]}\sum_{0<m\leq n, m\in Z-\Delta}
 [({n\over2}-m)k][({n\over2}-m)l]\eeq
\beq
C^{FR}(n\vert k,l)={\eps\over[k][l]}\sum_{0<m\leq n, m\in Z-\Delta}
 [({n\over2}-m)k][({n\over2}-m)l]_+\eeq
\beq
C^{RR}(n\vert k,l)={\eps\over[k][l]}\sum_{0<m\leq n, m\in Z-\Delta}
 [({n\over2}-m)k]_+[({n\over2}-m)l]_+.
\eeq
We thereby note that eq.\eq{Fbc} forms a closed algebra by itself and the
algebra \eq{FFbc} coincides with the $q$-${\rm Vir}^F$ commutation relation.
We can construct the OPE formula also for this ghost realization of
$q$-${\rm Vir}^F$ in the following way. Defining field corresponding to
$F^{bc}(n,k)$
\beq
F^{bc}(z;k)=\sum_n F^{bc}(n,k)z^{-n-2}=
{-1\over[k](q-q^{-1})z} \Bigl(
       q^{k(1-2\Delta)/2}b(zQ^{-1})c(zQ) -(q\ra q^{-1}) \Bigr) \label{Fbcz}
\eeq
and using
\beq
F^{bc}(z;k)b(w)={-1\over[k](q-q^{-1})w}\Bigl(
             Q^{2-2\Delta}{b(wQ^{-2})\over zQ-w}-(q\ra q^{-1}) \Bigr)
\eeq
\beq
F^{bc}(z;k)c(w)={1\over[k](q-q^{-1})w}\Bigl(
             Q^{-2\Delta}{c(wQ^2)\over zQ^{-1}-w}-(q\ra q^{-1}) \Bigr),
\eeq
we obtain the OPE corresponding to \eq{FFbc}
\[ \hskip -70pt
F^{bc}(z;k)F^{bc}(w;l)={\eps\over[k][l](q-q^{-1})^2zw}\sum_{\eps,\nu}
{\eps\nu(Q^\eps P^\nu)^{1-2\Delta}\over(zQ^{-\eps}-wP^\nu)(zQ^\eps-wP^{-\nu})}
\]
\beq  \hskip 20pt  + {1\over w(q-q^{-1})}\sum_{\eps,\eta=\pm1}
             {[\eps k+\eta l]\over[\eps k][\eta l]}
          {P^{-\eta}\over zQ^{-\eps}-wP^{\eta}} F^{bc}(wQ^\eps;\eps k+\eta l).
\eeq
This is same as \eq{opeFF} up to difference between central terms.


One may expect to construct a superalgebra for $q$-${\rm Vir}^{bc}$ or
$q$-${\rm Vir}^F$ in terms of $bc$ and $\beta\gamma$ ghosts. Unfortunately we
have not succeeded in constructing a desirable superalgebra which reduces to
the conventional Virasoro superalgebra. We here only present a basic closed
superalgebra, which might produce some kind of deformed Virasoro superalgebra
through recombination. Introducing $\beta\gamma$ fields with conformal
dimensions $\lambda$ in addition to \eq{Fbc} and \eq{Rbc}, we can verify that
\beq
F^{\beta\gamma}(n,k)={-1\over[k]}\sum_{m\in Z+\lambda}
                     [k({n\over2}-m)]:\beta_{n-m}\gamma_m:
\eeq
\beq
R^{\beta\gamma}(n,k)={1\over[k]}\sum_{m\in Z+\lambda}
                     [k({n\over2}-m)]_+:\beta_{n-m}\gamma_m:
\eeq
and
\beq
J^{\pm}(n,k)=\sum_{m\in Z-\Delta}q^{k(m-n/2)}:b_m\gamma_{n-m}:\pm
\sum_{m\in Z+\Delta}q^{k(m-n/2)}:c_m\beta_{n-m}:
\eeq
satisfy the following closed algebra
\beq \hskip -10pt
[F^{bc}(n,k),J^{\pm}(m,l)]={1\over2[k][1]_-}
\left( -q^{mk-nl\over2}J^\pm(n+m,l+k)+q^{-mk-nl\over2}J^\pm(n+m,l-k)\right)
\eeq
\beq\hskip -10pt
[F^{\beta\gamma}(n,k),J^{\pm}(m,l)]={1\over2[k][1]_-}
\left( q^{-mk+nl\over2}J^\pm(n+m,l+k)-q^{mk+nl\over2}J^\pm(n+m,l-k)\right)
\eeq
\beq \hskip -15pt
[R^{bc}(n,k),J^{\pm}(m,l)]={1\over2[k]}
\left( q^{mk-nl\over2}J^\mp(n+m,l+k)+q^{-mk-nl\over2}J^\mp(n+m,l-k)\right)
\eeq
\beq \hskip -10pt
[R^{\beta\gamma}(n,k),J^{\pm}(m,l)]={1\over2[k]}
\left( -q^{-mk+nl\over2}J^\mp(n+m,l+k)-q^{mk+nl\over2}J^\mp(n+m,l-k)\right)
\eeq
\[ \hskip -10pt
\{J^\pm(n,k),J^\pm(m,l)\}=\pm2[k-l]\left( q^{mk+nl\over2}R^{bc}(n+m,k-l)
+ q^{-{mk+nl\over2}}R^{\beta\gamma}(n+m,k-l) \right)\]\beq
\hskip
-15pt\pm(q-q^{-1}){[n(k-l)/2]\over[(k-l)/2]}[(k-l)f(\Delta)]\delta_{n+m,0}
\eeq
\[
\{J^\pm(n,k),J^\mp(m,l)\}=\pm2[k-l][1]_-\left(q^{mk+nl\over2}F^{bc}(n+m,k-l)
+ q^{-{mk+nl\over2}}F^{\beta\gamma}(n+m,k-l)\right )\]\beq
   \hskip -5pt\mp2{[n(k-l)/2]\over[(k-l)/2]}
              [(k-l)(\Delta-[\Delta]_G-{1\over2})]_+\delta_{n+m,0},
\eeq
where
\beq
f(\Delta)= \left\{
\begin{array}{ll}
   0,                         &\quad \mbox{for integral $\Delta$}  \\
   \Delta-[\Delta]_G-1/2,   &\quad \mbox{for rational $\Delta$}
\end{array}\right.
\eeq
and $[{\phantom C}]_G$ means the Gauss bracket.

\setcounter{equation}{0}
\section{Conclusion}
\indent

We have calculated the OPE formulae for three types of the deformed
super-Virasoro algebras and considered a construction of deformed Virasoro
algebra in terms of ghost Heisenberg oscillators. The most important problem
is that the deformed Virasoro algebra discussed here in $bc$ system has
different structure constants from those of $q$-${\rm Vir}^F$ or
$q$-${\rm Vir}^B$ and we do not know which deformed Virasoro algebra should be
shared by bosonic, fermionic and ghost realizations. If the
$q$-${\rm Vir}^B$/$q$-${\rm Vir}^F$ is a common algebra, we have to find its
realization in terms of $bc$/bosonic system. If $q$-${\rm Vir}^{bc}$, we must
find both bosonic and fermionic realizations. We may have to examine various
possibility including generalized algebras \cite{KS} or standing on a more
fundamental basis like in \cite{LP},\cite{DVA1},\cite{DVA2} to find an
appropriate unique algebra.

Although the determination of common structure constants is necessary for
supersymmetric argument and cancellation of anomalies etc., we can demonstrate
how the critical dimensions may depend on the deformation parameter $q$
supposing only fermionic and ghost realization of $q$-${\rm Vir}^F$.
The anomaly free equation for fermion-ghost $q$-${\rm Vir}^F$ operator
$L(n,k)=F(n,k)+F^{bc}(n,k)-a(k)\delta_{n,0}$, where $a(k)$ is a shift of
zero mode, leads us to a relation between $a(k)$ and the number of fermion
$D$;
\[ a(k)={-(2D+q^{k/2}+q^{-k/2})\over[k][k/2](q-q^{-1})^2}. \]
This allows that $D$ might vary depending on the value of $q$. Note that the
$q$-${\rm Vir}^F$ in $bc$ system certainly satisfies the Virasoro algebra
in $q\ra1$, however this limit is meaningless because there is no
corresponding field operator in $q=1$ as can be easily seen from \eq{Fbcz}.
While the $q\not=1$ case, we can expand $F^{bc}(n,k)$ into an infinite sum of
operators with various conformal weights, and we hence regard it as a kind of
$W_{1+\infty}$ algebra \cite{KS2}. Nevertheless to find a similar equation
as above is interesting in the context of deforming string theories.

All of the deformed algebras discussed here belong not to the quantum groups
but to the Lie algebras. If we want to construct a deformed string theory
identifying its worldsheet with quantum plane which is a basis of quantum
groups, we must deal with a field theory on quantum plane. To extract the
Fourier mode operators of the Virasoro type constraints from quantum plane,
we need a contour integral on quantum plane, however it has not been defined
yet. At present, it is natural to consider that our deformation might be a
profile projected from strings on quantum spacetime. The relationship between
quantum groups and our deformed Virasoro algebras should be clarified
in future study.

There are some unsolved problems suggested above, however we expect that
our analysis presented here may provide interesting and useful information
for integrable models, string and conformal field theories as well as
for application to quantum groups and $W$-infinity symmetries \cite{sato}.

\vspace{1cm}
\noindent
{\em Acknowledgements}

The author would like to thank T. Kobayashi for discussion and
N. Aizawa, S. Lukyanov and S. Odake for useful communication.

\newpage
\setcounter{equation}{0}
\appendix
\section{}
\indent

All commutation relations between generators and fields for
Chaichian-Presnajder's superalgebra are summarized.
\beq
\{G(r,k),\psi(z)\}=z^{r+1/2}q^{-k(z\del+r/2+1)}\Phi(z)
\eeq
\beq
[F(n,k),\psi(z)]=z^n{1\over[k]}[{k(z\del+{1\over2})+nk/2}]\psi(z)
\eeq
\beq
[G(r,k),\Phi(z)]=z^{r-1/2}q^{kz\del+k(r+1)/2}[z\del+r+{1\over2}]\psi(z)
\eeq
\beq
[B(n,k),\Phi(z)]=z^n\sum_{\eta=\pm1}{\eta\over2}
                  [(k+\eta)(z\del+1)+{n\over2}(k+2)]\Phi(z)
\eeq
\beq
[F(n,k),G(z;l)]={z^n\over[k](q-q^{-1})}q^{nl/2}\sum_{\eps=\pm1}\eps
                    q^{\eps k(z\del+3/2+n)/2}G(z;l+\eps k)
\eeq
\beq
[B(n,k),G(z;l)]={z^n\over2(q-q^{-1})}q^{-nl/2}\sum_{\eps,\eta=\pm1}\eta
      q^{\eps k(z\del+3/2+n)/2}q^{\eta(z\del+{3\over2}+2n)/2}
      G(z;\eps k-l+\eta)
\eeq
\[ \hskip -50pt
\{G(r,k),G(z;l)\}=C^G(r\vert l,k) z^{r-3/2}
              +2z^{r+1/2}q^{rl/2}q^{-k(z\del+r+2)/2} B(z;k-l)\]\beq
     \hskip 40pt  +z^{r+1/2}q^{-rl/2}q^{k(z\del+r+2)/2}\sum_\eps\eps
                 q^{\eps k(z\del+2r+2)/2}[k-l+\eps] F(z;k-l+\eps)
\eeq
\beq
[B(n,k),B(z;l)]=C^B(n\vert k,l) z^{n-2}
         + z^n \sum_{\eps,\eta}{\eta\over2}
        [{1\over2}(k+\eta)(z\del+2)+{n\over2}(k+\eps l+2\eta)]
        B(z;k+\eps l+\eta)
\eeq
\beq
[F(n,k),F(z;l)]=C^F(n\vert k,l) z^{n-2}
         + z^n \sum_{\eps}{[k+\eps l]\over[k][l]}
        [{\eps k\over2}(z\del+2)+{n\over2}(\eps k+l)] F(z;k+\eps l)
\eeq


%
\end{document}